\documentclass{article}

\usepackage{PRIMEarxiv}

\usepackage[utf8]{inputenc} 
\usepackage[T1]{fontenc}    
\usepackage{hyperref}       
\usepackage{url}            
\usepackage{booktabs}       
\usepackage{amsfonts}       
\usepackage{nicefrac}       
\usepackage{microtype}      
\usepackage{lipsum}
\usepackage{amsmath}
\usepackage{fancyhdr}       
\usepackage{graphicx}       
\graphicspath{{media/}}     

\title{An Enhanced Whale Optimization Algorithm with Log-Normal Distribution for Optimizing Coverage of Wireless Sensor Networks
\thanks{\textit{\underline{Citation}}: 
\textbf{Authors. Title. Pages.... DOI:000000/11111.}} 
}

\author{
  Junhao Wei, Yanzhao Gu, Ran Zhang, Wenxuan Zhu, Jinhong Song, Yapeng Wang, Xu Yang*, Ngai Cheong*\\
  Faculty of Applied Sciences \\
  Macao Polytechnic University \\
  Macao, China\\
  \texttt{\{p2312195,p2311998,p2512396,p2525981,p2525620,yapengwang,xuyang,ncheong\}@mpu.edu.mo}  
}

\begin{document}
\maketitle

\begin{abstract}
Wireless Sensor Networks (WSNs) are essential for monitoring and communication in complex environments, where coverage optimization directly affects performance and energy efficiency. However, traditional algorithms such as the Whale Optimization Algorithm (WOA) often suffer from limited exploration and premature convergence. To overcome these issues, this paper proposes an enhanced WOA which is called GLNWOA. GLNWOA integrates a log-normal distribution model into WOA to improve convergence dynamics and search diversity. GLNWOA employs a Good Nodes Set initialization for uniform population distribution, a Leader Cognitive Guidance Mechanism for efficient information sharing, and an Enhanced Spiral Updating Strategy to balance global exploration and local exploitation. Tests on benchmark functions verify its superior convergence accuracy and robustness. In WSN coverage optimization, deploying 25 nodes in a 60 m $\times$ 60 m area achieved a 99.0013\% coverage rate, outperforming AROA, WOA, HHO, ROA, and WOABAT by up to 15.5\%. These results demonstrate that GLNWOA offers fast convergence, high stability, and excellent optimization capability for intelligent network deployment.
\end{abstract}

\keywords{Wireless Sensor Networks \and Coverage Optimization \and Whale Optimization Algorithm \and Log-Normal Distribution.}

\section{Introduction}
Over the past few decades, optimization algorithms have played a crucial role in solving complex optimization problems in engineering, science, and intelligent systems. Traditional deterministic optimization methods, such as the gradient descent and Newton methods, often struggle with nonlinear, high-dimensional, non-differentiable, or multi-modal problems. These methods are prone to becoming trapped in local optima and exhibit strong dependence on the problem model and initial conditions. To overcome these limitations, researchers have successively proposed metaheuristic algorithms, which perform global searches by simulating biological evolution, animal behavior, or physical phenomena to approximate optimal solutions in complex search spaces. Since the 1990s, metaheuristic algorithms have developed rapidly. Representative algorithms include the Genetic Algorithm (GA) \cite{GA}, Particle Swarm Optimization \cite{PSO}, Ant Colony Optimization \cite{ACO}, Artificial Bee Colony \cite{ABC}, Grey Wolf Optimizer (GWO) \cite{GWO}, and Whale Optimization Algorithm (WOA) \cite{WOA}. However, despite their wide applicability, metaheuristic algorithms still face several challenges, such as the difficulty of balancing exploration and exploitation, parameter sensitivity, reduced population diversity in later iterations, and a tendency to fall into local optima. To address these issues, researchers have proposed numerous hybrid and improved variants. In 2021, P. Nandagopal incorporated the concept of the BAT algorithm into WOA and proposed WOABAT to determine the optimal positions of static WBANs \cite{WOABAT}. In 2022, Yu-Jun Zhang et al. developed the Logistic Map-Randomized Arithmetic Optimization Algorithm (LMRAOA), which combines multi-leader wandering search, random high-speed jumps, and adaptive lens opposition-based learning, significantly improving convergence accuracy and efficiency \cite{LMRAOA}. In 2023, J. Geng et al. proposed the Modified Adaptive Sparrow Search Algorithm (MASSA), integrating chaotic opposition-based learning, dynamic adaptive weighting, and a spiral search strategy to enhance convergence speed and global optimization performance \cite{MASSA}. In 2024, Wei et al. introduced an Improved Particle Swarm Optimization (IPSO) algorithm that combines Tent chaotic mapping, Levy flight, and adaptive t-distribution mechanisms to achieve faster convergence and higher precision \cite{IPSO}. In 2025, Gu et al. fused the Sine Cosine Algorithm (SCA) with WOA and proposed GWOA, which effectively enhances global exploration and prevents premature convergence \cite{GWOA}. Due to their gradient-free nature, simple structure, and ease of implementation, these algorithms have been widely applied in fields such as adjustment of hyperparameters in neural networks \cite{xgb}, path planning \cite{drrt} \cite{AHRRT}, feature selection, engineering design, and Wireless Sensor Networks (WSNs) \cite{251} \cite{252}.\par
In WSNs, coverage optimization is one of the key factors affecting network performance and energy efficiency \cite{nexus}. A WSN consists of a large number of sensor nodes capable of sensing, computation, and communication within a monitoring region. The core challenge lies in maximizing the coverage area with a limited number of nodes. Traditional coverage optimization methods, based on geometric analysis, partitioning strategies, or deterministic planning, often suffer from high computational cost and sub-optimal performance in complex or large-scale environments. With the development of metaheuristic algorithms, researchers have increasingly applied them to WSN coverage optimization—using PSO, GA, ACO, and GWO to optimize node positions or activation strategies—thereby achieving higher coverage and lower energy consumption.\par
However, despite the success of existing metaheuristics in WSN coverage optimization, they still exhibit certain shortcomings, such as insufficient global exploration, poor convergence accuracy, and difficulty in maintaining a balance between search speed and stability. To address these limitations, this paper proposes an enhanced Whale Optimization Algorithm based on the Log-Normal Distribution, termed GLNWOA. The algorithm introduces a log-normal perturbation mechanism into the original WOA framework to dynamically regulate the search step size and direction, thereby strengthening the balance between global exploration and local exploitation. Through this modification, GLNWOA effectively mitigates premature convergence and achieves faster convergence and higher coverage rates in WSN optimization problems.\par

\section{Whale Optimization Algorithm}
The Whale Optimization Algorithm (WOA) was proposed by Mirjalili et al. in 2016 as a metaheuristic optimization method inspired by the social hunting behavior of humpback whales \cite{WOA}. In WOA, two key foraging behaviors of whales (encircling prey and spiral updating position) are mathematically modeled to guide the population toward the global optimum.

\subsection{Encircling Prey}
Humpback whales are capable of identifying the position of prey and surrounding it during hunting. Since the exact location of the global optimum in the search space is unknown, WOA assumes that the current best candidate solution represents the target prey or is close to it. Once the best search agent is determined, other agents update their positions relative to it. This behavior can be described by the following equations:
\begin{equation}
    {D} = \left| {C} \cdot {X}^*(t) - {X}(t) \right|
    \label{eq1}
\end{equation}
\begin{equation}
    {X}(t+1)={X}^*(t)-{A}\cdot {D}
    \label{eq2}
\end{equation}
where $t$ denotes the current iteration; $A$ and $C$ are coefficient vectors; $X^*$ represents the position of the current best solution; and $X$ is the position vector of the whale. If a new position yields a better fitness value than $X^*$, the best position is updated accordingly.\par
The coefficient vectors $A$ and $C$ are defined as:
\begin{equation}
    {A}=2{a}\cdot {r}-{a}
    \label{eq3}
\end{equation}

\begin{equation}
    C=2\cdot {r}
    \label{eq5}
\end{equation}
where $r$ is a random vector within [0,1], and convergence factor $a$ decreases linearly from 2 to 0 over the course of iterations. \par
\begin{figure}[t!]
    \centering
    \includegraphics[width=0.8\textwidth]{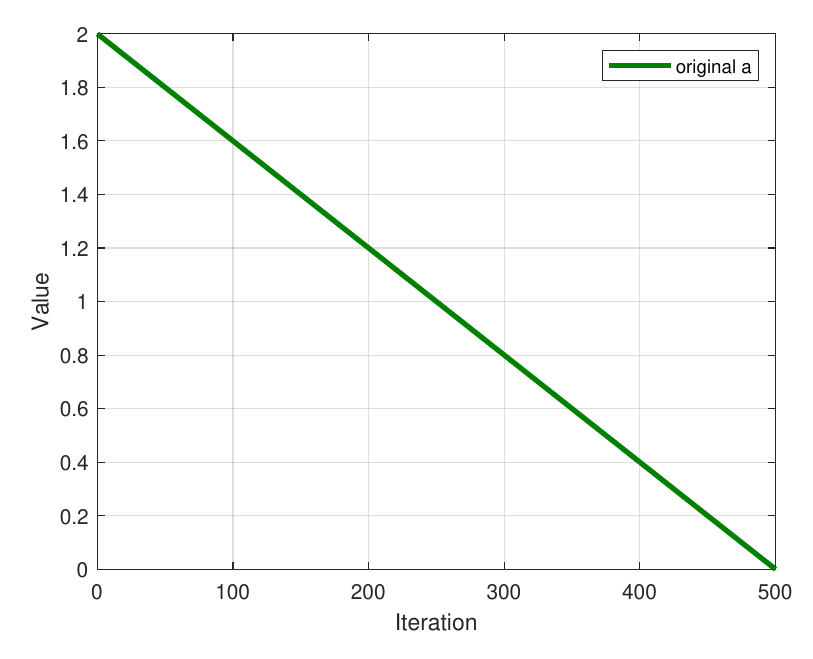}
    \caption{The variation of the convergence factor $a$ with iterations.}
    \label{origin_a}
\end{figure}

The convergence factor $a$ is updated as:
\begin{equation}
    {a}=2-2\cdot\frac tT
    \label{eq4}
\end{equation}
Fig.~\ref{origin_a} shows the variation process of the convergence factor $a$.

\subsection{Attacking Prey}
In addition to surrounding prey, humpback whales perform a bubble-net feeding maneuver to attack their prey. They create a spiral-shaped net of air bubbles while swimming upward around the prey at a depth of approximately 15 meters. This unique behavior is modeled in WOA through two strategies: shrinking encircling mechanism and spiral updating position.
\subsubsection{Shrinking encircling mechanism}
This process is realized by reducing the value of |A| in Eq.~\ref{eq3}, which gradually shrinks the encircling boundary. The position update is performed as described in Eq.~\ref{eq2}.\par
\subsubsection{Spiral updating position}
In this mechanism, the distance between the whale and the prey is first computed as:\par
\begin{equation}
    {D'}=|{X}^*(t)-{X}(t)|
    \label{eq7}
\end{equation}
The whale then moves along a spiral path around the prey using the following logarithmic spiral model:\par
\begin{equation}
    {X}(t+1)={X}^*(t)+{D'}\cdot e^{bl}\cdot\cos(2\pi l)
    \label{eq6}
\end{equation}
\begin{equation}
    l=(a_1-1)\cdot Rand+1
    \label{eq8}
\end{equation}
where $b$ is a constant defining the shape of the logarithmic spiral (typically $b$=1); $l$ is a random number in [-2,1]; and the parameter $a_1$ is linearly decreased as:\par
\begin{equation}
    a_1=-1-\frac tT
    \label{eq9}
\end{equation}

\subsection{Search-for-Prey}
When a whale moves beyond the prey's position, it randomly explores new directions in the search space to avoid local optima. This stochastic search behavior can be modeled as follows:\par
\begin{equation}
    {D}^{\prime\prime}=|{C}\cdot{X}_{rand}-{X}(t)|
    \label{eq11}
\end{equation}
\begin{equation}
    {X}(t+1)={X}_{rand}-{A}\cdot {D}''
    \label{eq12}
\end{equation}
where ${X}_{rand}$ is a random whale selected from the current population; ${A}$ and ${C}$ are described in Eq.~\ref{eq3} and Eq.~\ref{eq5}.

\subsection{Initialization}
In the original WOA, the population is initialized using a pseudo-random number approach:
\begin{equation}
    {X}_{i,j}=(ub-lb)\cdot Rand+lb
    \label{eq13}
\end{equation}
where ${X}_{i,j}$ denotes the position of the $i^{th}$ whale in the $j^{th}$ dimension; $ub$ and $lb$ represent the upper and lower bounds of the search space; $Rand$ is a random number uniformly distributed within [0,1].

\subsection{Advantages and Limitations of WOA}
The Whale Optimization Algorithm (WOA) features a relatively simple structure with few parameters, making it easy to understand and implement. By dynamically adjusting the convergence factor $a$, WOA emphasizes global exploration during the early search phase and local exploitation in the later phase, thereby preventing premature convergence to local optima. However, although WOA demonstrates a certain balance between exploration and exploitation, it often struggles with multimodal optimization problems. Specifically, WOA may fail to adequately explore the global search space and tends to converge prematurely to local optima, particularly when facing complex or high-dimensional landscapes. Moreover, for certain specific problems, WOA exhibits a relatively slow convergence rate, especially in the later iterations when population diversity decreases significantly, leading to stagnation. Additionally, WOA has difficulty maintaining an effective balance between exploration and exploitation. To address these limitations, this study proposes the GLNWOA. 

\section{The Proposed GLNWOA}
\subsection{Good Nodes Set Initialization}
\begin{figure}[t!]
    \centering
    \includegraphics[width=\textwidth]{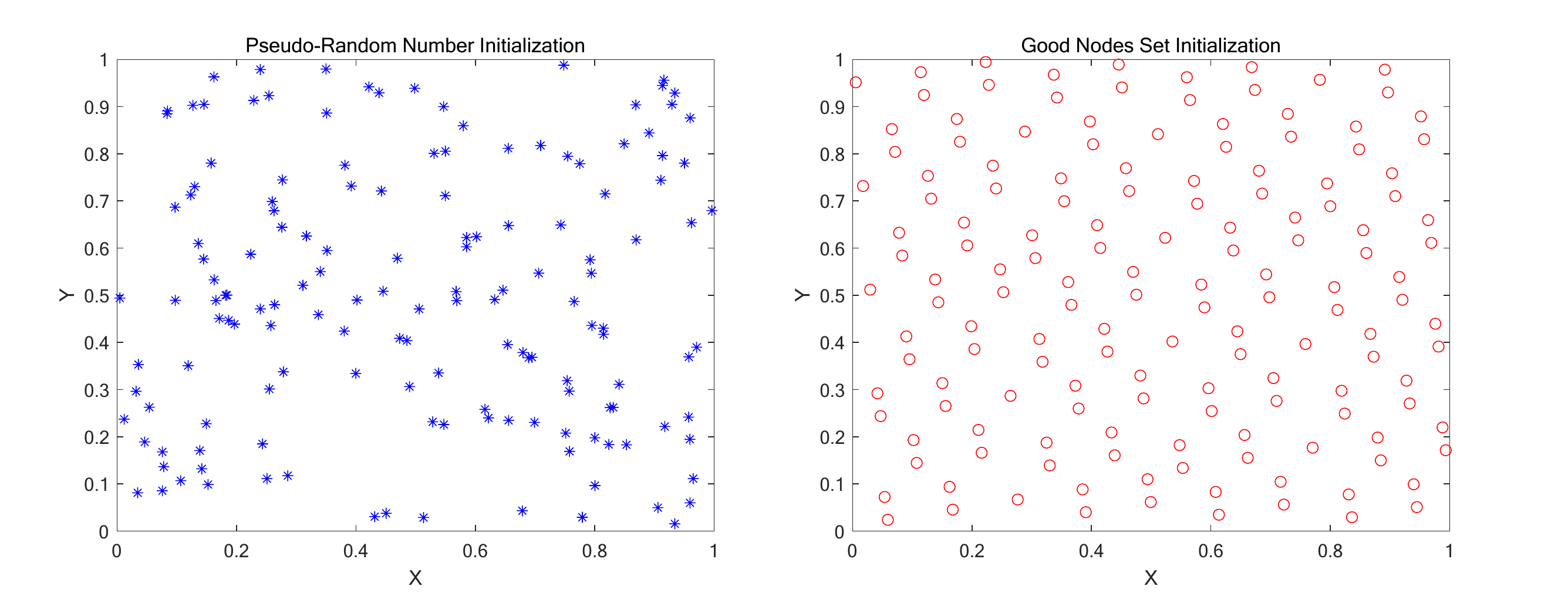}
    \caption{Comparison of different initialization methods. (a) Left: population generated by pseudo-random initialization with $N$=150; (b) Right: population generated by Good Nodes Set initialization with $N$=150} 
    \label{fig3}
\end{figure}

The traditional WOA algorithm employs pseudo-random number initialization to generate the population. Pseudo-random number initialization is straightforward and highly random; however, the individuals generated in this manner are often unevenly distributed across the search space. This can lead to clustering phenomena, resulting in poor population diversity, premature convergence, and low exploration efficiency during the search process. The left panel of Fig.~\ref{fig3} illustrates the population distribution generated by pseudo-random initialization when the population size is $N$=150, where clear clustering and vacant regions are observable. To address the shortcomings of pseudo-random initialization, various chaotic mapping initialization strategies have been proposed. Nevertheless, the Tent map is essentially still a pseudo-random distribution method based on chaotic sequences, and its uniformity cannot be rigorously guaranteed in high-dimensional spaces.\par
Therefore, this study adopts the Good Nodes Set (GNS) initialization method to further enhance the uniformity of the population distribution in the search space \cite{LSEWOA} \cite{MRBMO} \cite{TSWOA}. The concept of the Good Nodes Set was first proposed by the Chinese mathematician Hua Luogeng and provides a mathematical approach to generate uniformly distributed point sets in high-dimensional spaces. Its core idea is to construct sequences independent of dimensionality, ensuring that the projections of the point set in any dimension maintain good uniformity, thereby significantly improving the coverage and quality of the initial population.
Suppose $U^D$ represents the unit hypercube in a $D$-dimensional Euclidean space, and $r$$\in$$U^D$. Then, the Good Nodes Set  $P_r^M$ can be expressed as:\par
\begin{equation}
   P_r^M=\{p(k)=(\{kr\},\{kr^2\},\ldots,\{kr^D\})|k=1,2,\ldots,M\}
    \label{eq14}
\end{equation}
where $\{X\}$ denotes the fractional part of $X$; $M$ is the number of points; $r$ is a positive deviation parameter; and the constant $C(r,\varepsilon)$, which depends only on $r$ and $\varepsilon$, is a positive constant.\par
Assuming the $i^{th}$ dimension of the search space has upper and lower bounds  $x_{max}^i$ and $x_{min}^i$, respectively, the mapping of the Good Nodes Set to the actual search space is given by:\par
\begin{equation}
    {X}_k^i={X}_{min}^i+p(k)\cdot({X}_{max}^i-{X}_{min}^i)
    \label{eq15}
\end{equation}

\subsection{Leader Cognitive Guidance Mechanism}
In the original Whale Optimization Algorithm (WOA), the 'Search-for-Prey' mechanism updates positions by randomly selecting individuals to achieve global search. However, this mechanism overly relies on guidance from random individuals, resulting in unstable search directions, individual dispersion, and susceptibility to local optima. Moreover, random prey search lacks effective cognitive feedback in the later iterations, limiting the utilization of leader information for precise convergence. To enhance directional guidance and information fusion during the search phase, this study proposes the Leader Cognitive Guidance Mechanism, modeled as follows:
\begin{equation}
    {X}(t+1)=(1-\frac tT)^2 \cdot {X}^*(t)+|{X}_R(t)-{X}^*(t)|
     \label{eq16}
\end{equation}

\begin{equation}
    {X}_R(t)=\frac1N\sum_{i=1}^N{X}_i(t)
     \label{eq18}
\end{equation}
where $t$ is the current iteration, $T$ is the maximum number of iterations, $X$ represents the position of a whale individual, $X^*$ is the current best solution, $X_R$ denotes the mean position of all individuals (population average), and $N$ is the population size.\par
By introducing a nonlinear decay term of the leader's position and the deviation from the population mean, this mechanism allows individuals to adaptively adjust under leader cognition, forming a balance between global exploration and local exploitation. Compared with the original Search-for-Prey mechanism, the Leader Cognitive Guidance Mechanism maintains diversity in early iterations, avoids premature convergence, and enables more precise convergence control in later iterations, significantly improving search stability and global optimization performance.

\subsection{Dynamic Spiral Convergence Mechanism}
In the original WOA, the 'Encircling Prey' mechanism controls convergence around the leader via a linear contraction coefficient. However, this update is relatively simplistic; individuals lack dynamic perturbation and directional adjustment during convergence, which can lead to premature convergence or excessively fast convergence, reducing global search ability. To address this, the Dynamic Spiral Convergence Mechanism is introduced, modeled as follows \cite{LSWOA}:
\begin{equation}
	{X}(t+1)={X}^*(t)+h(s)\cdot|{A}\cdot {D}|
	 \label{eq19}
 \end{equation}
\begin{equation}
	 h(s)=e^{Z\cdot j}\cdot cos(2\pi j)
	 \label{eq19.5}
 \end{equation}
 \begin{equation}
	{D}=|{C}\cdot {X}^*(t)-{X}(t)|
	 \label{eq20}
 \end{equation}
 \begin{equation}
	j=2\cdot r-1
	 \label{eq21}
 \end{equation}
 \begin{equation}
	Z=e^{cos(\pi\cdot(1-\frac tT))}
	 \label{eq22}
\end{equation}
Where $A$ and $C$ are coefficient vectors, $h(s)$ is the spiral flight step, $j$ and $Z$ are spiral coefficients, $r$ is a random number in [0,1], $D$ is the distance between an individual and the current best solution, and $l$$\in$[-1,1] is a random spiral direction control factor.\par
Dynamic Spiral Convergence Mechanism introduces an exponential-cosine coupled spiral perturbation during the encircling phase, enabling individuals to converge around the leader in a dynamically rotating manner. The spiral radius gradually contracts via nonlinear temporal adjustment, preserving search diversity while enhancing attraction toward the optimum. Compared with the traditional Encircling Prey mechanism, Dynamic Spiral Convergence Mechanism effectively mitigates premature convergence and improves local exploitation and convergence precision in the middle and later iterations.

\subsection{Enhanced Spiral Updating Mechanism}
In the original WOA, the spiral updating mechanism simulates the whale's rotating movement around prey to update positions. However, this mechanism tends to fall into local optima in later iterations because the spiral radius decays monotonically and lacks dynamic perturbation, limiting population diversity. To overcome this, this study introduces log-normal distribution perturbation and proposes the Enhanced Spiral Updating Mechanism, formulated as:\par
\begin{equation}
    {X}(t+1)=\xi \cdot {X}^*(t)+{D'}\cdot e^{bl}\cdot\cos(2\pi l)
    \label{eq100}
\end{equation}
where $\xi$$~$$LogNormal(\mu,\sigma)$, and its probability density function is:\par
\begin{equation}
    f(\xi)=\frac{1}{\xi\sigma\sqrt{2\pi}}e^{-\frac{(\ln\xi-\mu)^2}{2\sigma^2}},\quad\xi>0    
    \label{eq101}
\end{equation}
The log-normal distribution has a right-skewed long-tail property. When $\mu =0$ and $\sigma  =0.5$, most values of $\xi$ concentrate near 1, but there is a small probability of generating large perturbations. This introduces stochastic diffusion while maintaining local convergence capability. Thus, the formula adds a log-normal random perturbation term to the original spiral update, increasing the probability of escaping local optima in later iterations. Compared with the original spiral mechanism, the Enhanced Spiral Updating Mechanism dynamically balances local exploitation and global exploration, enhancing search diversity and global optimization ability.

\subsection{Convergence Factor $a$}
In the original Whale Optimization Algorithm (WOA), the convergence factor $a$ controls the balance between global exploration and local exploitation, and its update follows a linear decreasing scheme:\par
\begin{equation}
    \vec{a}=2-2\cdot\frac tT
    \label{eq4}
\end{equation}
where $t$ is the current iteration count; $T$ is the maximum number of iterations.\par
\begin{figure}[htbp]
    \centering
    \includegraphics[width=0.8\textwidth]{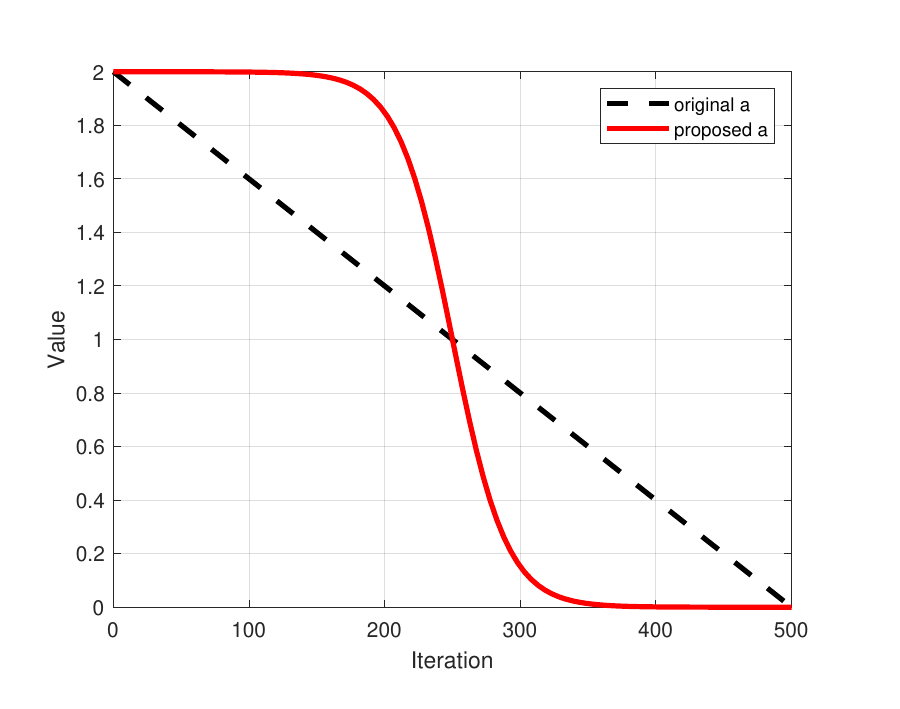}
    \caption{The original convergence factor $a$ and the proposed $a$}
    \label{compare_a}
\end{figure}

This linear update reduces $a$ uniformly from 2 to 0, meaning that the algorithm exhibits strong global search capability in the early stage and gradually shifts toward local exploitation in later iterations. However, this linear decrement is overly simplistic: it decreases too rapidly in the initial iterations, causing search agents to converge prematurely to local regions and weakening the sufficiency of global exploration; conversely, in the middle and later stages, the decrease is too slow, which can negatively affect convergence speed. To address this limitation, this study introduces a reversed Sigmoid function to achieve nonlinear dynamic adjustment of convergence factor $a$, as shown in Fig.~\ref{compare_a} the update formula as follows:\par
\begin{equation}
    a=2-\frac2{1+e^{-25(\frac tT-0.5)}}
    \label{eq27}
\end{equation}
where $t$ is the current iteration count; $T$ is the maximum number of iterations.\par
This formula maintains a relatively high value of convergence factor $a$ during the early iterations with a slow decline, decreases rapidly in the middle stage, and flattens in the later stage. Through this S-shaped variation, the algorithm can maintain sufficient global search coverage in the early stage, quickly focus near the optimal solution in the middle stage, and converge smoothly in the final stage. Compared with the linear decrement, the proposed Sigmoid-based nonlinear update strategy dynamically balances exploration and exploitation across different search stages, effectively preventing premature convergence and improving overall optimization efficiency \cite{RWOA}.

\section{Comparative Experiment}
The experimental environment for this study consisted of Windows 11 (64-bit), an Intel(R) Core(TM) i5-8300H CPU @ 2.30 GHz processor, 8 GB of RAM, and MATLAB R2023a as the simulation platform. To evaluate the optimization performance of the GLNWOA algorithm, a series of experiments were designed in which the algorithm was tested on selected benchmark functions, as listed in Table 1.
For comparison, the Harris Hawk Optimization algorithm (HHO) \cite{HHO}, Attraction-Repulsion Optimization Algorithm (AROA) \cite{AROA}, Remora Optimization Algorithm (ROA) \cite{ROA}, Whale Optimization Algorithm (WOA) \cite{WOA}, and WOABAT were tested on the same benchmark functions presented in Table~\ref{table1}. The parameter settings for all algorithms are summarized in Table~\ref{table2}. In all experiments, the number of iterations was set to $T$=500 and the population size to $N$=30. Each algorithm was independently run 30 times on each of the benchmark functions, and the mean, standard deviation, Wilcoxon test $p$-values, and Friedman test values were recorded for performance analysis.\par

\begin{table}[htbp]
    \centering
    \caption{Classical Benchmark Functions}
    \resizebox{\linewidth}{!}{
    \begin{tabular}{c c c c c}
    \hline
    Function & Function's Name & Type & Dimension & Best Value \\
    \hline
    F1 & Sphere & Uni-modal & 30 & 0 \\
    F2 & Schwefel's Problem 2.22 & Uni-modal & 30 & 0 \\
    F3 & Schwefel's Problem 2.21 & Uni-modal & 30 & 0 \\
    F4 & Generalized Rosenbrock's Function & Uni-modal & 30 & 0 \\
    F5 & Generalized Schwefel's Function & Multi-modal & 30 & -12569.5 \\
    F6 & Generalized Griewank's Function & Multi-modal & 30 & 0 \\
    F7 & Generalized Penalized Function 2 & Multi-modal & 30 & 0 \\
    F8 & Kowalik's Function & Multi-modal & 4 & 0.0003075 \\
    F9 & Six-Hump Camel-Back Function & Multi-modal & 2 & -1.0316 \\
    F10 & Branin Function & Multi-modal & 2 & 0.398 \\
    F11 & Goldstein-Price Function & Multi-modal & 2 & 3 \\
    F12 & Hartman's Function 1 & Multi-modal & 3 & -3.8628 \\
    F13 & Hartman's Function 2 & Multi-modal & 6 & -3.32 \\
    F14 & Shekel's Function 1 & Multi-modal & 4 & -10.1532 \\
    F15 & Shekel's Function 2 & Multi-modal & 4 & -10.4029 \\
    F16 & Shekel's Function 3 & Multi-modal & 4 & -10.5364 \\
    \hline
    \label{table1}
    \end{tabular}}
\end{table}

\begin{table}[htbp]
    \centering
    \caption{Parameter settings for different metaheuristic algorithms}
    \begin{tabular}{ccc}
        \hline
        Algorithm & Parameter & Value \\
        \hline
		HHO & Threshold & 0.5 \\ \hline
		AROA & Attraction factor $c$ & 0.95 \\ 
		& Local search scaling factor 1 & 0.15 \\ 
		& Local search scaling factor 2 & 0.6 \\ 
		& Attraction probability 1 & 0.2 \\ 
		& Local search probability & 0.8 \\ 
		& Expansion factor & 0.4 \\ 
		& Local search threshold 1 & 0.9 \\
		& Local search threshold 2 & 0.85 \\ 
		& Local search threshold 3 & 0.9 \\ \hline
		ROA & $C$ & 0.1 \\
        & Velocity Factor $V$ & 2 decreasing to 0\\
        & Convergence Factor $a$ & -1 decreasing to -2\\\hline
		WOA & Convergence factor $a$ & 2 decrease to 0 \\
        & Spiral factor $b$ & 1 \\ \hline
        WOABAT & Convergence factor $a$ & 2 decrease to 0 \\
        & Spiral factor $b$ & 1 \\ \hline
        GLNWOA & Convergence factor $a$ & 2 decrease to 0 \\
        & Spiral factor $k$ & 1 \\ 
        \hline
        \label{table2}
    \end{tabular}
\end{table}

\begin{figure}[htbp]
    \centering
    \includegraphics[width=1\textwidth]{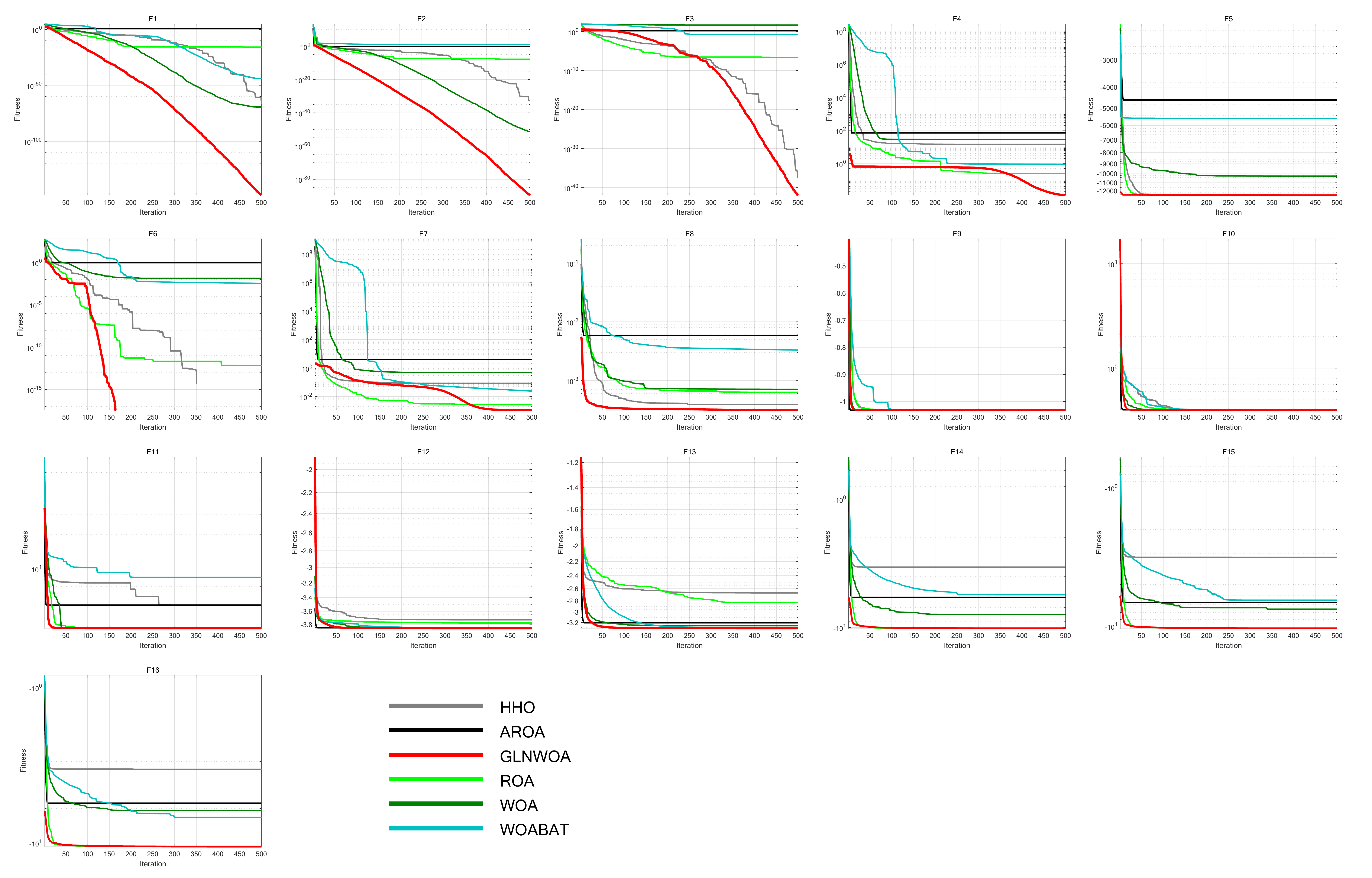}
    \caption{Iteration curves of the algorithms}
    \label{differ}
\end{figure}

The experimental results are presented in Fig.~\ref{differ} and Table~\ref{tab:benchmark_results} and Table~\ref{nonpara}. The experimental results demonstrate that, when solving the benchmark functions, GLNWOA outperforms all other algorithms in terms of both mean and standard deviation. During the optimization process, GLNWOA consistently locates the optimal solutions quickly, exhibiting faster convergence and higher solution accuracy. These findings indicate that GLNWOA possesses strong adaptability and robustness when handling different types of optimization problems.

\begin{table*}[htbp]
\centering
\caption{Comparison results of parametric tests of  different algorithms}
\begin{tabular}{cccccccc}
\hline
Function & Metrics & HHO & AROA & GLNWOA & ROA & WOA & WOABAT \\
\hline
F1  & Ave & 5.6138E-67 & 5.1848E+00 & 8.7935E-149 & 1.7294E-16 & 5.0072E-70 & 9.2331E-45 \\
    & Std & 3.0748E-66 & 5.4365E+00 & 4.4139E-148 & 4.7451E-16 & 2.7425E-69 & 2.7933E-44 \\
F2  & Ave & 3.1718E-33 & 6.6773E-01 & 2.6184E-90 & 1.5712E-08 & 3.9855E-52 & 9.3333E+00 \\
    & Std & 1.7373E-32 & 2.1424E-01 & 1.0060E-89 & 8.1028E-08 & 9.7127E-52 & 1.1725E+01 \\
F3  & Ave & 3.5192E-38 & 1.6395E+00 & 1.1087E-42 & 2.1825E-07 & 4.8051E+01 & 1.7211E-01 \\
    & Std & 1.4679E-37 & 8.2230E-01 & 3.7075E-42 & 7.9372E-07 & 2.6584E+01 & 8.9922E-01 \\
F4  & Ave & 1.4230E+01 & 6.9347E+01 & 1.1971E-02 & 2.4710E-01 & 2.7968E+01 & 8.8632E-01 \\
    & Std & 1.4099E+01 & 3.0179E+01 & 1.0957E-02 & 4.2977E-01 & 4.1052E-01 & 4.8484E+00 \\
F5  & Ave & -1.2569E+04 & -4.5802E+03 & -1.2569E+04 & -1.2569E+04 & -1.0265E+04 & -5.5815E+03 \\
    & Std & 1.6987E-01 & 7.4079E+02 & 8.0838E-03 & 7.5861E-01 & 1.8430E+03 & 1.7142E+02 \\
F6  & Ave & 0.0000E+00 & 9.8196E-01 & 0.0000E+00 & 6.9541E-13 & 1.4388E-02 & 3.5097E-03 \\
    & Std & 0.0000E+00 & 1.2575E-01 & 0.0000E+00 & 2.8476E-12 & 5.6889E-02 & 1.9223E-02 \\
F7  & Ave & 8.4465E-02 & 4.0414E+00 & 1.1193E-03 & 2.6626E-03 & 4.8050E-01 & 2.3939E-02 \\
    & Std & 1.5346E-01 & 6.2054E-01 & 3.3581E-03 & 4.5619E-03 & 2.2867E-01 & 4.5134E-02 \\
F8  & Ave & 3.8732E-04 & 5.8160E-03 & 3.1235E-04 & 6.2905E-04 & 7.0488E-04 & 3.2956E-03 \\
    & Std & 1.3041E-04 & 7.9289E-03 & 1.8585E-05 & 4.3662E-04 & 2.9061E-04 & 6.2693E-03 \\
F9  & Ave & -1.0316E+00 & -1.0316E+00 & -1.0316E+00 & -1.0316E+00 & -1.0316E+00 & -1.0316E+00 \\
    & Std & 1.6411E-07 & 2.2185E-05 & 1.0807E-15 & 8.1950E-05 & 7.7058E-10 & 1.3927E-12 \\
F10 & Ave & 3.9794E-01 & 3.9992E-01 & 3.9789E-01 & 3.9862E-01 & 3.9789E-01 & 3.9789E-01 \\
    & Std & 6.9583E-05 & 6.5126E-03 & 1.4986E-14 & 1.6868E-03 & 1.6016E-05 & 1.5801E-10 \\
F11 & Ave & 4.8004E+00 & 4.8008E+00 & 3.0000E+00 & 3.0011E+00 & 3.0001E+00 & 8.4000E+00 \\
    & Std & 6.8501E+00 & 6.8499E+00 & 3.6308E-05 & 1.8420E-03 & 2.3472E-04 & 1.0985E+01 \\
F12 & Ave & -3.7301E+00 & -3.8548E+00 & -3.8628E+00 & -3.7775E+00 & -3.8562E+00 & -3.8628E+00 \\
    & Std & 1.9638E-01 & 1.4218E-02 & 5.5784E-07 & 8.7375E-02 & 1.2420E-02 & 6.1555E-07 \\
F13 & Ave & -2.6670E+00 & -3.2037E+00 & -3.3133E+00 & -2.8293E+00 & -3.2623E+00 & -3.2977E+00 \\
    & Std & 5.0612E-01 & 1.1148E-01 & 3.3334E-02 & 2.7664E-01 & 9.2061E-02 & 4.9346E-02 \\
F14 & Ave & -3.3906E+00 & -5.8296E+00 & -1.0153E+01 & -1.0150E+01 & -7.9315E+00 & -5.5650E+00 \\
    & Std & 1.4288E+00 & 3.3008E+00 & 1.3818E-10 & 5.5414E-03 & 2.7864E+00 & 1.5555E+00 \\
F15 & Ave & -3.1854E+00 & -6.7576E+00 & -1.0403E+01 & -1.0399E+01 & -7.5548E+00 & -6.5051E+00 \\
    & Std & 1.8900E+00 & 3.4626E+00 & 7.9705E-11 & 1.0028E-02 & 3.1644E+00 & 2.3907E+00 \\
F16 & Ave & -3.3593E+00 & -5.5407E+00 & -1.0536E+01 & -1.0534E+01 & -6.1785E+00 & -6.8407E+00 \\
    & Std & 1.5277E+00 & 3.1380E+00 & 8.4017E-11 & 5.6573E-03 & 3.2827E+00 & 2.9696E+00 \\
\hline
\end{tabular}
\label{tab:benchmark_results}
\end{table*}

\begin{table}[htbp]
    \centering
    \caption{Comparison results of non-parametric tests of different algorithms}
    \begin{tabular}{cccc}
        \hline
        Algorithm & Rank & Average Friedman Value & (+/=/-) \\
        \hline
        HHO  & 5 & 3.8302  & (15/1/0) \\
        AROA  & 6 & 5.0771   & (16/0/0) \\
        GLNWOA & 1 & 1.2750   & - \\
        ROA  & 4 & 3.7740  & (16/0/0) \\
        WOA  & 3 & 3.7615  & (16/0/0) \\
        WOABAT & 2 & 3.2823   & (16/0/0) \\
        \hline
    \end{tabular}
    \label{nonpara}
\end{table}

\section{Simulation}
Wireless Sensor Networks (WSNs) consist of a large number of sensor nodes equipped with sensing, computation, and communication capabilities. They have wide applications in environmental monitoring, target tracking, disaster warning, and military reconnaissance. Coverage optimization is one of the core challenges in WSN design, aiming to deploy or adjust sensor nodes in such a way that the monitoring area is maximally covered with the minimum number of nodes, thereby improving network sensing capability and energy efficiency. Traditional WSN coverage optimization often relies on deterministic algorithms (e.g., geometric planning, Voronoi partitioning, or gradient-based methods). However, these approaches generally require strict mathematical modeling, high continuity or convexity of the objective function, and are prone to becoming trapped in local optima in high-dimensional, complex search spaces. With the development of intelligent optimization theory, metaheuristic algorithms—characterized by derivative-free operation, global search capability, and strong robustness—have become effective tools for solving nonlinear, multimodal, and multi-constraint optimization problems.\par
In coverage optimization, the position coordinates of each sensor node can be treated as decision variables, and the collective positions of all nodes form the search space. The coverage rate ($CR$) serves as the fitness function (objective function), and the algorithm iteratively updates node positions to maximize the coverage. To verify the effectiveness and superiority of the proposed GLNWOA in WSN coverage optimization, this section discusses the coverage model construction, parameter and experimental setup, comparison algorithms, and experimental results.

\subsection{Coverage Model}
In this study, a two-dimensional rectangular monitoring area $\phi$ is considered, with the square region having a side length of $L$=60m and an area of $L$$\times$$L$. Within this monitoring area, $S$=25 mobile sensor nodes are randomly deployed, with the node set defined as:\par
\begin{equation}
    S=\{s_1,s_2,s_3,...,s_S\}
    \label{102}
\end{equation}
All sensor nodes have the same sensing radius $R_s$=8.39m and communication radius $R_c \geq 2R_s$. The sensing model adopts the Boolean Sensing Model. Assuming the monitoring area is divided into $m \times n$ regular grid points with a spacing of $d_L$=1m, the coordinates of each grid point $P_k(x_k,y_k)$ and its Euclidean distance to node $s_i(x_i,y_i)$ are defined as:\par
\begin{equation}
    d(s_i,P_k)=\sqrt{(x_i-x_k)^2+(y_i-y_k)^2}
    \label{103}
\end{equation}
A grid point $P_k$ is considered covered if the distance to node $s_i$ satisfies the sensing range condition $d(s_i,P_k)\leq R_s$. Otherwise, it is uncovered. The coverage determination function is expressed as:
\begin{equation}
    C(s_i,P_k)=
    \begin{cases}
        1, & d(s_i,P_k)\leq R_s \\
        0, & d(s_i,P_k)>R_s
    \end{cases}
    \label{104}
\end{equation}\par
Considering all nodes, the overall coverage of a grid point $P_k$ is given by:
\begin{equation}
    C(S,P_k)=1-\prod_{i=1}^S(1-C(s_i,P_k))
    \label{105}
\end{equation}
The coverage rate ($CR$) is defined as the proportion of grid points covered by at least one node relative to the total number of grid points:
\begin{equation}
    CR=\frac{\sum_{k=1}^{m\times n}C(S,P_k)}{m\times n}
    \label{106}
\end{equation}
where $CR$$\in$[0,1]. \par
$CR$ reflects the overall coverage capability of the node distribution in the monitoring area and serves as the primary metric for evaluating algorithm performance.

\subsection{Parameters and Environment}
The experimental environment for this study consisted of Windows 11 (64-bit), an Intel(R) Core(TM) i5-8300H CPU @ 2.30 GHz processor, 8 GB of RAM, and MATLAB R2023a as the simulation platform. To ensure reproducibility and fairness, all algorithms were tested under the same initial conditions and search parameters. The main parameter settings are summarized in Table~\ref{tab:sim_parameters}.\par
For all algorithms, the objective function was the coverage rate ($CR$) defined in Eq.~\ref{106}, which iteratively optimizes the positions of sensor nodes to maximize the overall coverage of the monitoring area. To reduce the influence of randomness, each experiment was independently run 20 times, and the mean value was taken as the final result.
\begin{table}[htbp]
\centering
\caption{Simulation parameter settings}
\begin{tabular}{ccc}
\hline
\textbf{Parameter} & \textbf{Symbol} & \textbf{Value} \\
\hline
Monitoring area & $L\times L$ & 60 m$\times$60 m \\
Grid interval & $d_L$ & 1 m \\
Number of sensor nodes & $S$ & 25 \\
Sensing radius & $R_s$ & 8.35 m \\
\hline
\end{tabular}
\label{tab:sim_parameters}
\end{table}

\subsection{Comparative Algorithms}
To comprehensively evaluate the optimization performance of GLNWOA, five representative metaheuristic algorithms were selected for comparison: Harris Hawk Optimization algorithm (HHO), Attraction-Repulsion Optimization Algorithm (AROA), Remora Optimization Algorithm (ROA), Whale Optimization Algorithm (WOA), and WOABAT. The parameter settings for these algorithms are provided in Table~\ref{table2}. In all experiments, the number of iterations was set to $T$=500 and the population size to $N$=30.

\subsection{Results}
\begin{figure}[htbp]
    \centering
    \includegraphics[width=0.7\textwidth]{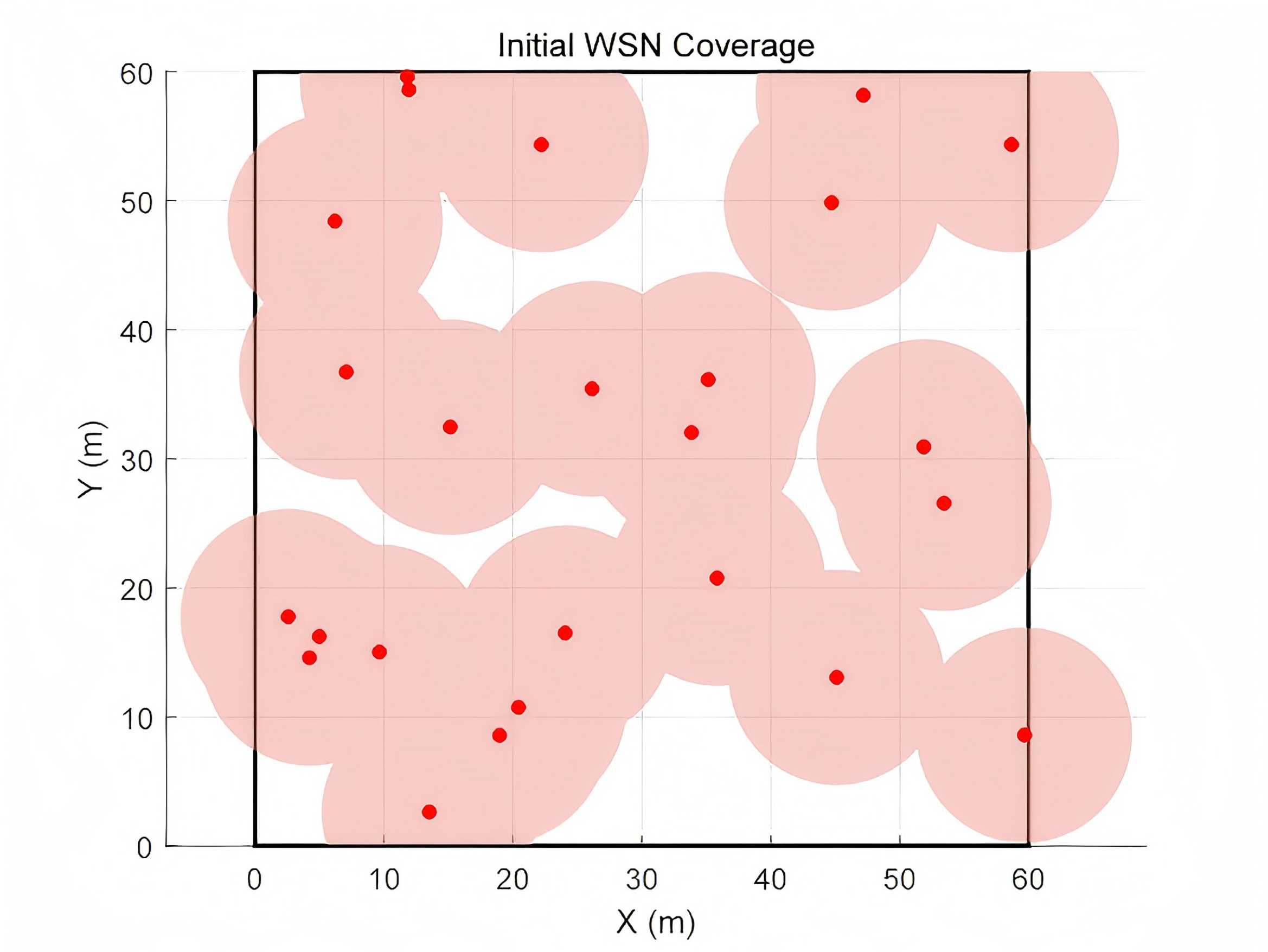}
    \caption{Random distribution of wireless sensor networks.}
    \label{ini_wsn}
\end{figure}

\begin{figure}[htbp]
    \centering
    \includegraphics[width=1\textwidth]{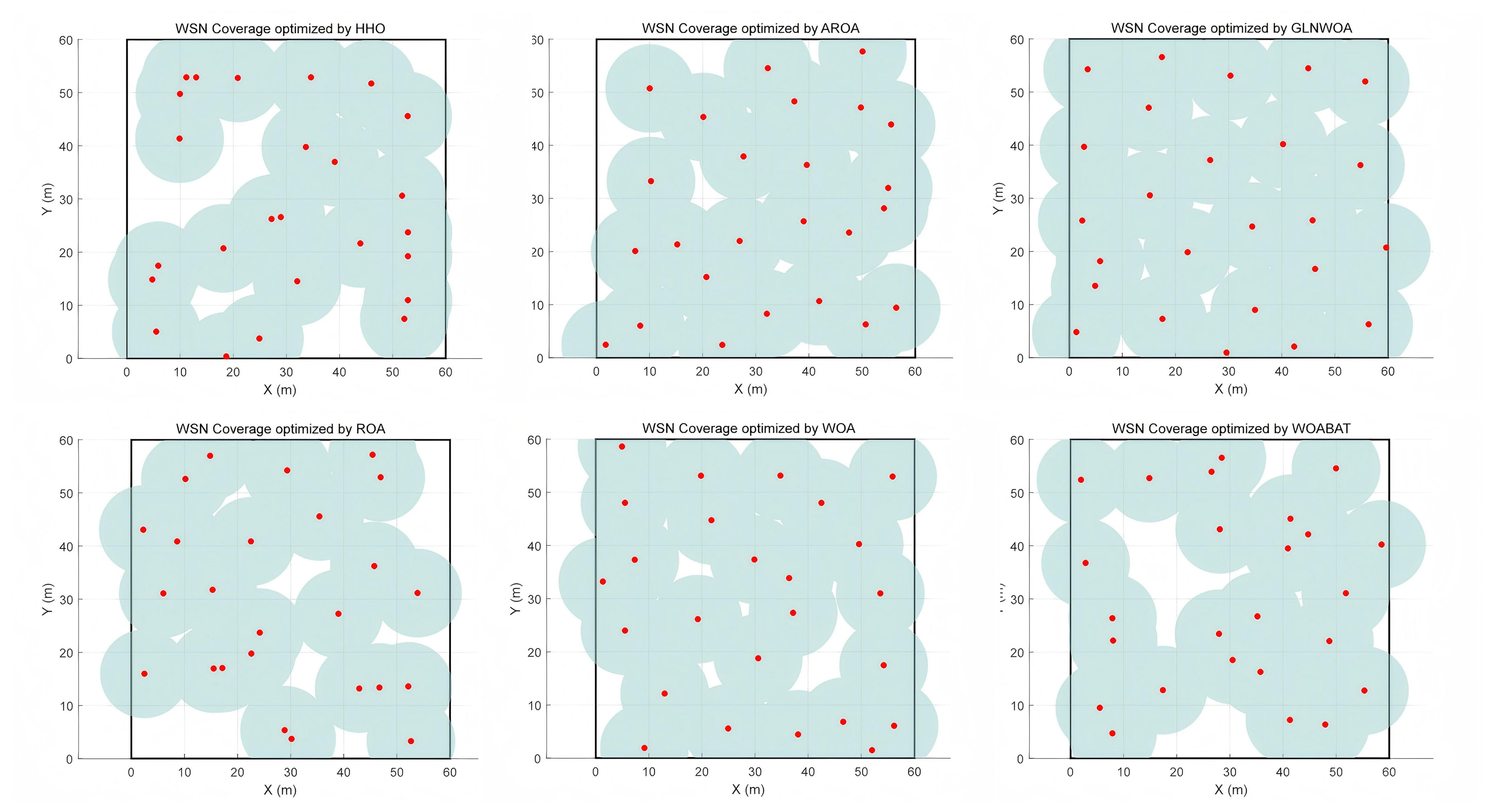}
    \caption{Experimental results of different algorithms optimizing the coverage of Wireless Sensor Networks.}
    \label{differ_wsn}
\end{figure}

\begin{figure}[htbp]
    \centering
    \includegraphics[width=0.9\textwidth]{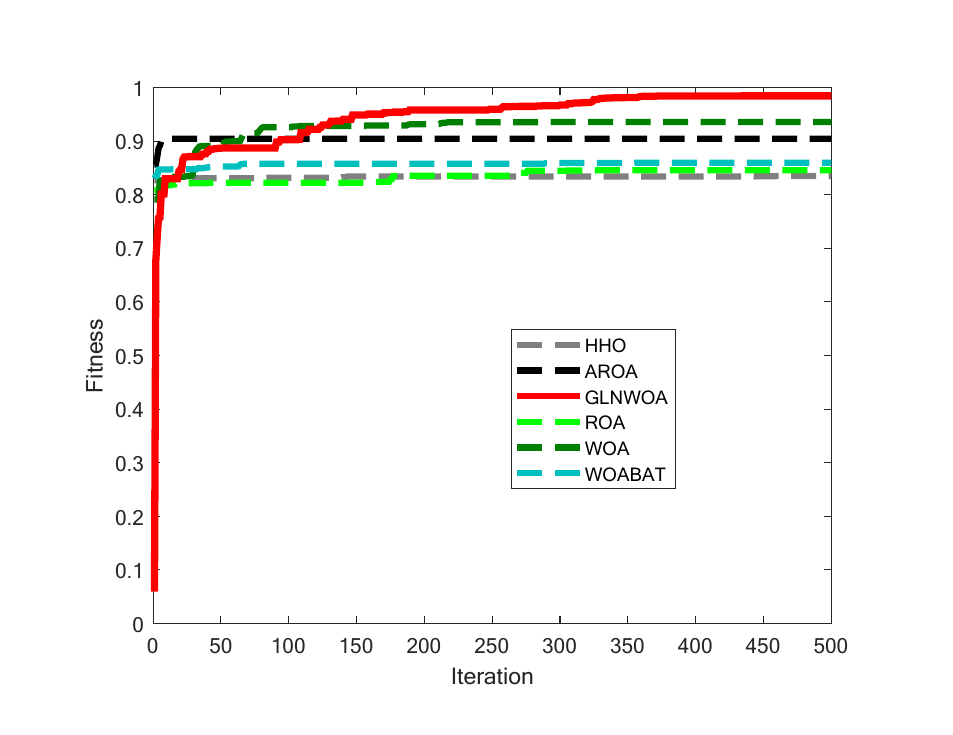}
    \caption{Iteration curves of different algorithms optimizing the coverage of Wireless Sensor Networks.}
    \label{iter_wsn}
\end{figure}

Fig.~\ref{ini_wsn} is a random distribution of wireless sensor networks. Fig.~\ref{differ_wsn} shows the WSN coverage maps optimized by different algorithms, while Fig.~\ref{iter_wsn} illustrates the convergence curves of these algorithms during the WSN coverage optimization process. Table~\ref{tab:coverage_comparison} presents the final coverage rates achieved by the six algorithms under the same parameter settings. It can be observed that the proposed GLNWOA achieves a significant improvement in coverage. Specifically, under identical experimental conditions, GLNWOA attained an average coverage rate of 99.0013\%, clearly outperforming the other comparison algorithms. Compared with the original WOA, this represents an increase of approximately 5.8\%, and compared with the relatively strong-performing AROA algorithm, an improvement of about 8.5\% is observed. These results indicate that GLNWOA exhibits stronger global search capability and higher coverage efficiency in optimizing the distribution of sensor nodes in wireless sensor networks.

\begin{table}[htbp]
\centering
\caption{Comparison of coverage rate ($CR$) among different algorithms}
\begin{tabular}{cc}
\hline
\textbf{Algorithm} & \textbf{Coverage Rate ($CR$)} \\
\hline
HHO & 83.4722\% \\
AROA & 90.4596\% \\
ROA & 84.5472\% \\
WOA & 93.5770\% \\
WOABAT & 85.9446\% \\
\textbf{GLNWOA (proposed)} & \textbf{99.0013\%} \\
\hline
\end{tabular}
\label{tab:coverage_comparison}
\end{table}

\section{Conclusion}
This paper addresses several limitations of the Whale Optimization Algorithm (WOA) in complex optimization problems, such as susceptibility to local optima, slow convergence in later stages, and insufficient search diversity, by proposing the GLNWOA. The proposed algorithm achieves a dynamic balance between global exploration and local exploitation through a multi-mechanism integration, and its superior performance is validated both theoretically and through simulation experiments.\par
In benchmark function experiments, GLNWOA outperformed WOA, AROA, HHO, ROA, and WOABAT in terms of convergence accuracy, speed, and stability, demonstrating strong robustness and cross-problem adaptability. In wireless sensor network (WSN) coverage optimization experiments, GLNWOA achieved a coverage rate of 99.0013\% with 25 nodes in a 60m $\times$ 60m monitoring area, significantly surpassing the other five algorithms. This confirms its exceptional performance in nonlinear and multi-modal optimization scenarios. Experimental results indicate that GLNWOA can rapidly identify global or near-global optimal node distributions, enabling near-complete coverage with fewer sensor nodes, thereby effectively enhancing network energy efficiency and monitoring capability. In summary, GLNWOA not only theoretically improves the search mechanism of WOA but also demonstrates strong engineering practicality and scalability in applications.

\section*{Acknowledgments}
The supports provided by Macao Polytechnic University (MPU Grant no: RP/FCA-03/2022; RP/FCA-06/2022) and Macao Science and Technology Development Fund (FDCT Grant no: 0044/2023/ITP2) enabled us to conduct data collection, analysis, and interpretation, as well as cover expenses related to research materials and participant recruitment. MPU and FDCT investment in our work have significantly contributed to the quality and impact of our research findings.

\bibliographystyle{unsrt}  
\bibliography{references}

\end{document}